\documentclass[showpacs,prd,superscriptaddress]{revtex4}
\usepackage{amsmath}
\usepackage{amssymb}
\usepackage{graphicx}
\usepackage{color}
\usepackage{dcolumn}

\begin{document}
\newcolumntype{.}{D{.}{.}{-1}}

\title{
Star-plus-wormhole systems with two interacting scalar fields
}

\author{Vladimir Dzhunushaliev}
\email[{\it Email:}]{v.dzhunushaliev@gmail.com}
\affiliation{
Institute of Experimental and Theoretical Physics,
Al-Farabi Kazakh National University, Almaty 050040, Kazakhstan
}
\affiliation{
Department of Theoretical and Nuclear Physics,
Al-Farabi Kazakh National University, Almaty 050040, Kazakhstan
}
\affiliation{
Institute of Physicotechnical Problems and Material Science of the NAS
of the Kyrgyz Republic, 265a, Chui Street, Bishkek, 720071,  Kyrgyz Republic
}
\author{Vladimir Folomeev}
\email[{\it Email:}]{vfolomeev@mail.ru}
\affiliation{
Institute of Physicotechnical Problems and Material Science of the NAS
of the Kyrgyz Republic, 265a, Chui Street, Bishkek, 720071,  Kyrgyz Republic
}
\author{Ajnur Urazalina}
\affiliation{
Institute of Experimental and Theoretical Physics,
Al-Farabi Kazakh National University, Almaty 050040, Kazakhstan
}

\begin{abstract}
We study static, spherically symmetric mixed configurations with a nontrivial
(wormhole) spacetime topology provided by the presence of two interacting ghost scalar fields.
Wormhole is assumed to be filled by a perfect relativistic neutron fluid modeled by a polytropic equation of state.
For such mixed configurations, we find regular, asymptotically flat
general relativistic solutions. It is shown that
the maximum of the fluid density is always shifted from the center,
and  the resulting configurations represent, in general, double-throat systems.
\end{abstract}


\maketitle

\section{Introduction}

In recent years interest in obtaining wormhole-type solutions
has increased appreciably,
primarily because of the discovery  of the accelerated expansion of the
present Universe. At the present time  it is widely believed that such an acceleration is due to the presence
of some special form of matter -- dark energy~(DE)~\cite{AmenTsu2010}. One of the key
features of the latter is its ability to violate various energy conditions. In the most extreme case,
DE is modeled by matter violating the null energy condition, when the effective pressure $p$ of matter filling the Universe
is negative and larger (modulus) than its energy density $\varepsilon$, i.e., $p< -\varepsilon$.
Such matter is called exotic.

In modeling the accelerated expansion of the Universe, it is usually assumed that  DE is distributed  homogeneously and isotropically on the largest scales.
This, however, does not exclude a possibility
that DE might cluster on relatively small scales comparable to sizes of galaxies or even of separate stars.
In the latter case, the literature in the field offers compact objects composed of DE (the so-called dark
energy stars),  which are modeled by some form of matter possessing the properties of DE \cite{de_stars}.
Such objects may have
both a trivial and a nontrivial topology of spacetime. In the latter case, to construct localized solutions, one needs to invoke some form of
exotic matter  which permits
compact configurations to have a nontrivial (wormholelike) topology.
In the simplest case
one can employ the so-called ghost (or phantom) scalar fields, which may be
massless~\cite{wh_massless} or possess a potential energy~\cite{wh_quart} (other examples of phantom field wormholes can be found, e.g., in
Refs.~\cite{wh_confs}, and for a  general overview
on the subject of Lorentzian wormholes, see the book~\cite{Visser}).

Another possibility is to consider the case where exotic matter is clustered in the presence of ordinary (baryonic) matter.
Such  mixed compact configurations, with a nontrivial topology provided by using ghost scalar fields, have been studied
in Refs.~\cite{mix_syst,Dzhunushaliev:2013lna,Dzhunushaliev:2014mza,Aringazin:2014rva}.
In that case, a wormhole is assumed to be filled by ordinary (neutron) matter, and the resulting systems
possess properties both of wormholes and of ordinary stars: on the one hand,
the topology of spacetime is nontrivial, and on the other hand, there is a visible surface created by a neutron fluid.
Due to the presence of the nontrivial topology, the physical properties (masses, sizes, distributions of matter) of such configurations
may differ considerably from those of systems with a trivial topology (for example, ordinary neutron stars).

The possibility to construct objects which acquire new properties in the presence of  a nontrivial topology
motivates one to study mixed systems supported by other sources of
exotic matter. In the simplest case one can modify the systems
of Refs.~\cite{mix_syst,Dzhunushaliev:2013lna,Dzhunushaliev:2014mza,Aringazin:2014rva}
``one scalar field plus ordinary matter'' by adding to them one more scalar field interacting with the first one.
Systems with two scalar fields are themselves well known in quantum field theory, where they are used to obtain solitary wave solutions~\cite{rajaraman}.
When a gravitational field is present,
such systems have also been repeatedly considered in the cosmological and astrophysical contexts (see, for example,
in Refs.~\cite{2_fields_syst}, as well as
at the beginning of Sec.~\ref{statem_prob}).
The presence of a second scalar field allows one to get new interesting solutions, not shared by one-field systems
(for example, the cosmological
quintom models containing usual and ghost scalar fields and possessing the properties both of quintessence and of
phantom models~\cite{AmenTsu2010}). We will show below that the inclusion in our mixed system of the second field  will  allow us to
obtain configurations that possess new physical properties and differ substantially from one-field systems.

Consistent with this,
in the present paper we consider mixed astrophysical systems consisting of a wormhole supported by two interacting ghost scalar fields
and filled by ordinary matter.
Our goal is to clarify the question of how the presence of such a wormhole
changes the distribution of the ordinary matter and influences the physical characteristics of the resulting objects.

The paper is organized as follows.
In Sec.~\ref{statem_prob} we
present the statement of the problem
and derive the corresponding general-relativistic equations
for the mixed systems under consideration.
In Sec.~\ref{num_calc} we solve these equations numerically
for different amount of ordinary matter and compare
the systems under consideration with configurations studied earlier.
Finally, in Sec.~\ref{conclusion} we summarize the results obtained.

\section{Statement of the problem}
\label{statem_prob}

As pointed out in the Introduction, here we study a mixed gravitating system consisting of
two interacting ghost scalar fields $\varphi, \chi$ and a perfect fluid.
The Lagrangian for this system can be presented in the form
\begin{equation}
\label{lagran_wh_star_poten_mix}
L=-\frac{c^4}{16\pi G}R-\left[\frac{1}{2}\partial_{\mu}\varphi\partial^{\mu}\varphi +
\frac{1}{2}\partial_{\mu}\chi\partial^{\mu}\chi-V(\varphi,\chi)\right]+L_m,
\end{equation}
with the curvature scalar $R$ and Newton's constant $G$.
Here $L_m$ is the Lagrangian of the perfect isotropic fluid
(where isotropic means that the radial and the tangential pressure
of the fluid agree),
which has the form $L_m=p$.

We choose the scalar field potential energy
$V(\varphi,\chi)$ as
\begin{equation}
\label{potph}
V(\varphi,\chi)=\frac{\lambda_1}{4}(\varphi^2-m_1^2)^2+\frac{\lambda_2}{4}(\chi^2-m_2^2)^2+\varphi^2
\chi^2-V_0.
\end{equation}
Here  $m_1$ and $m_2$ are the  masses of the scalar fields,
$\lambda_1, \lambda_2$
are the coupling constants, and $V_0$ is some constant whose value is chosen from the statement of the problem.
(Note that one of these free parameters can always be eliminated by the corresponding rescaling.)
Using this potential, we obtained earlier a number of solutions which can be employed both in describing astrophysical objects and
when considering cosmological problems. Namely, we have shown that:
(a) for the four-dimensional case, there exist regular
spherically and cylindrically
symmetric solutions
 \cite{2_fields_our,Dzhunushaliev:2007cs}, and also
cosmological solutions
both for usual fields (i.e., for the fields having the usual sign in front of the kinetic energy term)
and for ghost scalar fields~\cite{Dzhunushaliev:2006xh,Folomeev:2007uw};
(b) for the higher-dimensional cases,
there exist thick brane  solutions
 supported by usual and ghost scalar fields~\cite{2_fields_brane}.

An important feature of this potential is the presence of two local minima at  $\chi=0, \varphi=\pm~m_1$
to which the solutions tend asymptotically at spatial infinity.
These local minima correspond to two vacua. For  wormhole-type systems, the solution starts in one of these vacua as
$r\to - \infty$ and returns to it back as $r\to + \infty$~\cite{Dzhunushaliev:2007cs} ($r$ is the spatial coordinate).
Such solutions are called nontopological~\cite{rajaraman}, in contrast to
solitonlike solutions with one scalar field
when the solution starts in one of the vacua and goes to the other.
The latter solutions occurring  only in the presence of two or more vacua are called topological.
In the present paper we will discuss only the nontopological solutions.

\subsection{Field equations}

To describe the mixed equilibrium system under consideration, let us employ  the polar Gaussian coordinates,
in which the metric has the form
\begin{equation}
\label{metric_gauss}
ds^2=e^{\nu}(dx^0)^2- dr^2-R^2 d\Omega^2,
\end{equation}
where $\nu$ and $R$ are functions of $r$ only,
and $x^0=c\, t$ is the time coordinate.

The total energy-momentum tensor can be obtained from the Lagrangian~\eqref{lagran_wh_star_poten_mix} in the following form:
 \begin{equation}
\label{emt_wh_star_poten_mix}
T_i^k=(\varepsilon+p)u_i u^k-\delta_i^k p-\partial_{i}\varphi\partial^{k}\varphi-\partial_{i}\chi\partial^{k}\chi
-\delta_i^k\left[-\frac{1}{2}\partial_{\mu}\varphi\partial^{\mu}\varphi-\frac{1}{2}\partial_{\mu}\chi\partial^{\mu}\chi-V(\varphi,\chi)\right],
\end{equation}
where $\varepsilon$ and $p$ are the energy density
and the pressure of the fluid, and $u^i$ is the four-velocity.
By using~\eqref{emt_wh_star_poten_mix},  the
$(_0^0)$, $(_1^1)$, and $(_2^2)$ components of the Einstein equations with metric~\eqref{metric_gauss} take the form
\begin{eqnarray}
\label{Einstein-00_poten}
&&-\left[2\frac{R^{\prime\prime}}{R}+\left(\frac{R^\prime}{R}\right)^2\right]+\frac{1}{R^2}
=\frac{8\pi G}{c^4} T_0^0=
\frac{8\pi G}{c^4}\left[ \varepsilon-\frac{1}{2}\left(\varphi^{\prime 2}+\chi^{\prime 2}\right)-V(\varphi,\chi)\right],
 \\
\label{Einstein-11_poten}
&&-\frac{R^\prime}{R}\left(\frac{R^\prime}{R}+\nu^\prime\right)+\frac{1}{R^2}
=\frac{8\pi G}{c^4} T_1^1=
\frac{8\pi G}{c^4}\left[- p+\frac{1}{2}\left(\varphi^{\prime 2}+\chi^{\prime 2}\right)-V(\varphi,\chi)\right],
\\
\label{Einstein-22_poten}
&&\frac{R^{\prime\prime}}{R}+\frac{1}{2}\frac{R^\prime}{R}\nu^\prime+
\frac{1}{2}\nu^{\prime\prime}+\frac{1}{4}\nu^{\prime 2}
=-\frac{8\pi G}{c^4} T_2^2=
\frac{8\pi G}{c^4}\left[ p+\frac{1}{2}\left(\varphi^{\prime 2}+\chi^{\prime 2}\right)+V(\varphi,\chi)\right],
\end{eqnarray}
where the prime denotes  differentiation with respect to $r$.

The equations for the scalar fields $\varphi,\chi$ result from the Lagrangian \eqref{lagran_wh_star_poten_mix} in the general form as
\begin{equation}
\label{field_eq_gen}
\frac{1}{\sqrt{-g}}\frac{\partial}{\partial
x^\mu}\left[\sqrt{-g}\,\, g^{\mu\nu} \frac{\partial
(\varphi,\chi)}{\partial x^\nu}\right]=-\frac{\partial V}{\partial
(\varphi,\chi)}.
\end{equation}
In the metric \eqref{metric_gauss}, these equations yield
\begin{eqnarray}
\label{sf_eq_phi}
\varphi^{\prime\prime}+\left(\frac{1}{2}\nu^\prime+2\frac{R^\prime}{R}\right)\varphi^\prime=
\frac{d V}{d \varphi},\\
\label{sf_eq_chi}
\chi^{\prime\prime}+\left(\frac{1}{2}\nu^\prime+2\frac{R^\prime}{R}\right)\chi^\prime=
\frac{d V}{d \chi}.
\end{eqnarray}

Finally, the hydrostatic equation for the fluid can be obtained from the law of the
conservation of energy and momentum,
$T^k_{i;k}=0$. The $i=1$ component of this equation gives
\begin{equation}
\label{conserv_1}
\frac{d T^1_1}{d r}+
\frac{1}{2}\left(T_1^1-T_0^0\right)\nu^\prime+2\frac{R^\prime}{R}\left[T_1^1-\frac{1}{2}\left(T^2_2+T^3_3\right)\right]=0.
\end{equation}
Taking the components of the energy-momentum tensor from Eqs.~\eqref{Einstein-00_poten}-\eqref{Einstein-22_poten},
and also taking into account that $T_3^3=T_2^2$,
from \eqref{conserv_1} we have
\begin{equation}
\label{conserv_2}
\frac{d p}{d r}=-\frac{1}{2}(\varepsilon+p)\frac{d\nu}{d r}.
\end{equation}

The mixed system under investigation can be regarded as follows. Let initially we have a wormhole geometry
provided by two interacting ghost scalar fields. In the absence of ordinary matter (fluid), such systems have been considered in~\cite{Dzhunushaliev:2007cs}.
By adding a fluid to the system, here we consider the influence that such a wormhole has on the spatial distribution
of the fluid and on the physical characteristics (masses, sizes) of the resulting mixed systems.

As the fluid one can choose any type of matter used in modeling astrophysical objects (for instance, stars). One of the simplest ways
to describe star's matter is to employ a polytropic fluid. The latter is applied both in describing nonrelativistic objects
(Newtonian stars, see, e.g.,~\cite{Zeld}) and when considering relativistic systems,
including those with a relativistic neutron fluid in a strong gravitational field.
In doing so, a polytropic equation of state (EOS) can adequately represent  more realistic EOSs used in modeling neutron stars (see, e.g., \cite{n_eos}).

For our purpose, we choose the following relativistic EOS:
\begin{equation}
\label{eqs_NS_WH}
p=K \rho_{b}^{1+1/n}, \quad \varepsilon = \rho_b c^2 +n p,
\end{equation}
with the constant $K=k c^2 (n_{b}^{(ch)} m_b)^{1-\gamma}$,
the polytropic index $n=1/(\gamma-1)$,
and $\rho_b=n_{b} m_b$ denotes the rest-mass density
of the neutron fluid.
Here
$n_{b}$ is the baryon number density,
$n_{b}^{(ch)}$ is a characteristic value of $n_{b}$,
$m_b$ is the baryon mass,
and $k$ and $\gamma$ are parameters
whose values depend on the properties of the neutron matter.
In particular, here we choose
$m_b=1.66 \times 10^{-24}\, \text{g}$,
$n_{b}^{(ch)} = 0.1\, \text{fm}^{-3}$,
$k=0.1$, and $\gamma=2$~\cite{Salg1994}.

With the EOS in the form of \eqref{eqs_NS_WH},
one can integrate Eq.~\eqref{conserv_2}. To do this, it is convenient to introduce
the new variable $\theta$ \cite{Zeld},
\begin{equation}
\label{theta_def}
\rho_b=\rho_{b c} \theta^n~,
\end{equation}
where $\rho_{b c}$ is some characteristic density of the neutron fluid.
For the mixed configurations with an isotropic fluid considered by us earlier in Refs.~\cite{mix_syst,Dzhunushaliev:2013lna}
$\rho_{b c}$ corresponds to the central (maximum) density
at the wormhole throat. However, for systems with an anisotropic fluid a situation is possible where the maximum density of the neutron fluid is
shifted away from the center~\cite{Aringazin:2014rva}. Then $\rho_{b c}$ no longer plays the role of maximum density,
but it is just some characteristic value for the configuration under consideration.
As we will see below, here we deal with such a situation even in the case of an isotropic fluid.

Making use of expression \eqref{theta_def},
we have
from Eq.~\eqref{conserv_2}
\begin{equation}
\label{conserv_3}
2\sigma(n+1)\frac{d\theta}{d r}=
-\left[1+\sigma(n+1) \theta\right]\frac{d\nu}{dr},
\end{equation}
where $\sigma=K \rho_{b c}^{1/n}/c^2$ is the relativistic parameter~\cite{Tooper2}.
Integrating this equation, one can find
\begin{equation}
\label{theta_analit}
\theta=\frac{1}{\sigma (n+1)}\Big\{\left[1+\sigma (n+1)\theta_c\right]e^{(\nu_c-\nu)/2}-1\Big\},
\end{equation}
where $e^{\nu_c}$ is the value of $e^{\nu}$ at the center.
The integration constant $\nu_c$ is fixed
by requiring that the spacetime is asymptotically flat,
i.e., $e^{\nu}=1$ at infinity. In turn, the arbitrary constant
$\theta_c$, contained in the boundary conditions~\eqref{bound_stat},
corresponds to the central value of the function $\theta$.

For the numerical calculations, it is convenient to rewrite
the obtained equations in terms of dimensionless variables.
Since below we will seek solutions with zero central values of the scalar-field derivatives,
 $\varphi^\prime(0), \chi^\prime(0)=0$, and a nonzero central value of the potential energy,
 $V(\varphi(0),\chi(0))\neq 0$, we normalize the characteristic size of the system $L$  with respect to this value.
Namely, let us introduce dimensionless variables:
\begin{equation}
\label{dimless_var}
\xi=\frac{r}{L}, \quad \Sigma=\frac{R}{L},
\quad \tilde\varphi(\xi),\tilde\chi(\xi),\mu_{1,2}=\frac{\sqrt{8\pi G}}{c^2}\,\varphi(r),\chi(r),m_{1,2},
\quad \text{where} \quad L=\frac{c^2}{\sqrt{8\pi G |V(\varphi(0),\chi(0))|}}.
\end{equation}
In such variables Eqs.~\eqref{Einstein-00_poten}-\eqref{Einstein-22_poten}, \eqref{sf_eq_phi}, \eqref{sf_eq_chi} take the form
\begin{eqnarray}
\label{Einstein-00_dmls}
&&-\left[2\frac{\Sigma^{\prime\prime}}{\Sigma}+\left(\frac{\Sigma^\prime}{\Sigma}\right)^2\right]+\frac{1}{\Sigma^2}
=B  (1+\sigma n \theta) \theta^n
-\frac{1}{2}\left(\tilde\varphi^{\prime 2}+\tilde\chi^{\prime 2}\right)-\tilde{V},
 \\
\label{Einstein-11_dmls}
&&-\frac{\Sigma^\prime}{\Sigma}\left(\frac{\Sigma^\prime}{\Sigma}+\nu^\prime\right)+\frac{1}{\Sigma^2}
=-B \sigma \theta^{n+1}
+\frac{1}{2}\left(\tilde\varphi^{\prime 2}+\tilde\chi^{\prime 2}\right)-\tilde{V},
\\
\label{Einstein-22_dmls}
&&\frac{\Sigma^{\prime\prime}}{\Sigma}+\frac{1}{2}\frac{\Sigma^\prime}{\Sigma}\nu^\prime+
\frac{1}{2}\nu^{\prime\prime}+\frac{1}{4}\nu^{\prime 2}
=
B \sigma \theta^{n+1}
+\frac{1}{2}\left(\tilde\varphi^{\prime 2}+\tilde\chi^{\prime 2}\right)+\tilde{V},
\\
\label{sf_phi_dmls}
&&\tilde\varphi^{\prime\prime}+\left(\frac{1}{2}\nu^\prime+2\frac{\Sigma^\prime}{\Sigma}\right)\tilde\varphi^\prime=
\tilde\varphi\left[2 \tilde\chi^2+\Lambda_1\left(\tilde\varphi^2-\mu_1^2\right)\right],
\\
\label{sf_chi_dmls}
&&\tilde\chi^{\prime\prime}+\left(\frac{1}{2}\nu^\prime+2\frac{\Sigma^\prime}{\Sigma}\right)\tilde\chi^\prime=
\tilde\chi\left[2 \tilde\varphi^2+\Lambda_2\left(\tilde\chi^2-\mu_2^2\right)\right],
\end{eqnarray}
where the dimensionless potential energy is
\begin{equation}
\label{potph_dmls}
\tilde{V}=\frac{\Lambda_1}{4}(\tilde\varphi^2-\mu_1^2)^2+\frac{\Lambda_2}{4}(\tilde\chi^2-\mu_2^2)^2+\tilde\varphi^2
\tilde\chi^2-\tilde{V}_0.
\end{equation}
Here
$B=(\rho_{b c} c^2)/|V(\varphi(0),\chi(0))|$ is the dimensionless ratio of the characteristic fluid energy density to that
of the scalar field at the center; the dimensionless coupling constants $\Lambda_{1,2}=\left(c^4/8\pi G \sqrt{|V(\varphi(0),\chi(0))|}\right)^2 \lambda_{1,2}$, and
$\tilde{V}_0=V_0/|V(\varphi(0),\chi(0))|$.

Note that, using $L$ from \eqref{dimless_var}, the expression for
$B$ can be recast in the following form: $B=8\pi G \rho_{b c} (L/c)^2$.
This expression will be used below in performing numerical calculations.
In the case of $B=0$ we deal with a system consisting
of a pure scalar field configuration with no ordinary matter~\cite{Dzhunushaliev:2007cs}.

\subsection{Boundary conditions}

Here we consider neutron-star-plus-wormhole configurations
that are asymptotically flat and
symmetric under $\xi \to -\xi$.
The metric function $\Sigma(\xi)$ may be considered
as a circumferential radial coordinate.
Asymptotic flatness requires that $\Sigma(\xi) \to |\xi|$ for large $|\xi|$.
Because of the assumed symmetry,
the center of the configurations at
$\xi=0$ should correspond to an extremum
of $\Sigma(\xi)$, i.e.,~$\Sigma'(0)=0$.
If $\Sigma(\xi)$ has a minimum at $\xi=0$, then $\xi=0$ corresponds to the throat
of the wormhole.
If, on the other hand, $\Sigma(\xi)$ has a local maximum at $\xi=0$,
then $\xi=0$ corresponds to an equator.
In that case, the wormhole will have a double throat
surrounding a belly (for examples of double-throat systems,
see, e.g., Refs.~\cite{doub_wh,Dzhunushaliev:2014mza}).

Consistent with this,  Eqs.~\eqref{Einstein-00_dmls}-\eqref{sf_chi_dmls}, together with the expression~\eqref{theta_analit}, will be solved
for given parameters of the fluid $\sigma$, $n$, and $B$,
subject to the boundary conditions at the center of the configuration $\xi=0$,
\begin{equation}
\label{bound_stat}
\theta(0) = \theta_c, \quad \Sigma(0)=\Sigma_c, \quad \Sigma^\prime (0)=0,
\quad \nu(0)=\nu_c, \quad  \nu^\prime (0)=0,\quad
\tilde\varphi(0) =\tilde\varphi_c, \quad \tilde\varphi^\prime (0)=0,
\quad
\tilde\chi(0) =\tilde\chi_c, \quad \tilde\chi^\prime (0)=0,
\end{equation}
where $\theta_c,\Sigma_c,\tilde\varphi_c$, and $\tilde\chi_c$ are some constants.

To find the location of the throat,
expand the metric function
$\Sigma$ in the neighborhood of the center as
$$\Sigma\approx \Sigma_c+1/2\, \Sigma_2 \xi^2.$$
Then, using Eqs.~\eqref{Einstein-00_dmls} and \eqref{Einstein-11_dmls}, we find
the relations
\begin{equation}
\label{bound_coef}
\Sigma_c=\frac{1}{\sqrt{-B \sigma\theta_c^{n+1} -\tilde V(\tilde\varphi_c,\tilde\chi_c)}}, \quad
\Sigma_2=-\frac{\Sigma_c}{2}B\theta_c^n\left[1+(n+1)\sigma \theta_c\right].
\end{equation}
It is seen from these expressions that (i) to obtain physically reasonable solutions, one needs to choose the free parameters of the system  in such a way that
the radicand in the first formula  be positive, and
(ii) at $\theta_c>0$ (as it should be for ordinary matter) $\Sigma_2$ is negative,
corresponding to a solution with an equator
surrounded by a double throat.

\section{Numerical results}
\label{num_calc}

In Ref.~\cite{Dzhunushaliev:2007cs} we studied wormhole systems
supported by ghost scalar fields with the potential \eqref{potph}.
Here we modify that system by adding to it ordinary matter in the form of a polytropic fluid.
Our aim is to study such mixed configurations, in particular,
with respect to the effect that
the presence of the nontrivial topology has on the distribution of the neutron matter along the radius of the system.

We will solve the system of equations ~\eqref{Einstein-00_dmls}-\eqref{sf_chi_dmls} together with~\eqref{theta_analit}
numerically using the boundary conditions \eqref{bound_stat} and \eqref{bound_coef}.
Since a characteristic property of a gravitating polytropic fluid is the presence of an edge
where its pressure (density) goes to zero,
 the configurations under consideration can be subdivided into two regions:
 (i) the internal one, where both the scalar fields and the fluid are present;
 (ii) the external one, where only the scalar fields are present.
Correspondingly, the solutions in the external region are obtained by using Eqs.~\eqref{Einstein-00_dmls}-\eqref{sf_chi_dmls},
in which $\theta$ is set to zero.

The internal solutions must be matched with the external ones at the boundary of the fluid,
$\xi=\xi_b$, by equating the corresponding values of the functions $\tilde\varphi,\tilde\chi$, $\Sigma$, $\nu$
and their derivatives.
The boundary of the fluid $\xi_b$ is defined by $p(\xi_b)=0$.
Knowledge of the asymptotic solutions in turn allows one
to determine the value of the integration constant $\nu_c$ at the center,
proceeding from the requirement
of asymptotic flatness of the external solutions.

As in the case without ordinary matter studied in Ref.~\cite{Dzhunushaliev:2007cs},
the system of equations ~\eqref{Einstein-00_dmls}-\eqref{sf_chi_dmls} has regular solutions
(satisfying the necessary boundary conditions) only for certain values
of the coupling constants $\Lambda_1, \Lambda_2$ and of the masses
of the scalar fields $\mu_1, \mu_2$.
As a result, the problem reduces to a search for {\it eigenvalues} of the parameters $\mu_1, \mu_2$  and  for the
corresponding {\it eigenfunctions} $\nu, \Sigma, \tilde\varphi$, and $\tilde\chi$ of the nonlinear system of
differential equations~\eqref{Einstein-00_dmls}-\eqref{sf_chi_dmls}.
We will seek the specified eigenvalues by using the shooting method. A step-by-step description of the  procedure for finding solutions
can be found in Refs.~\cite{Dzhunushaliev:2007cs,Dzhunushaliev:2006xh}.

\begin{figure}[t]
\begin{minipage}[t]{.49\linewidth}
  \begin{center}
  \includegraphics[width=7.3cm]{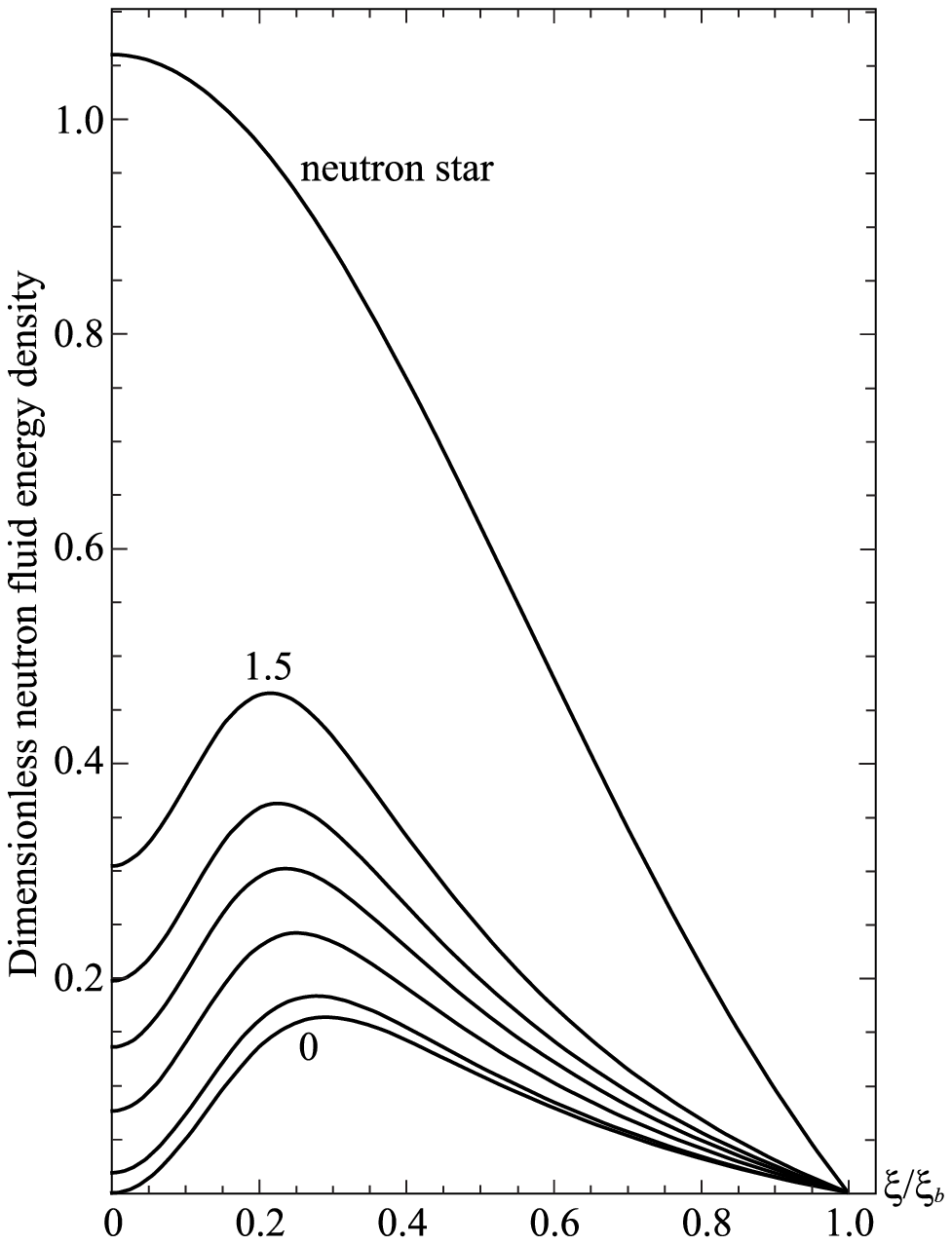}
  \end{center}
\end{minipage}\hfill
\begin{minipage}[t]{.49\linewidth}
  \begin{center}
  \includegraphics[width=7cm]{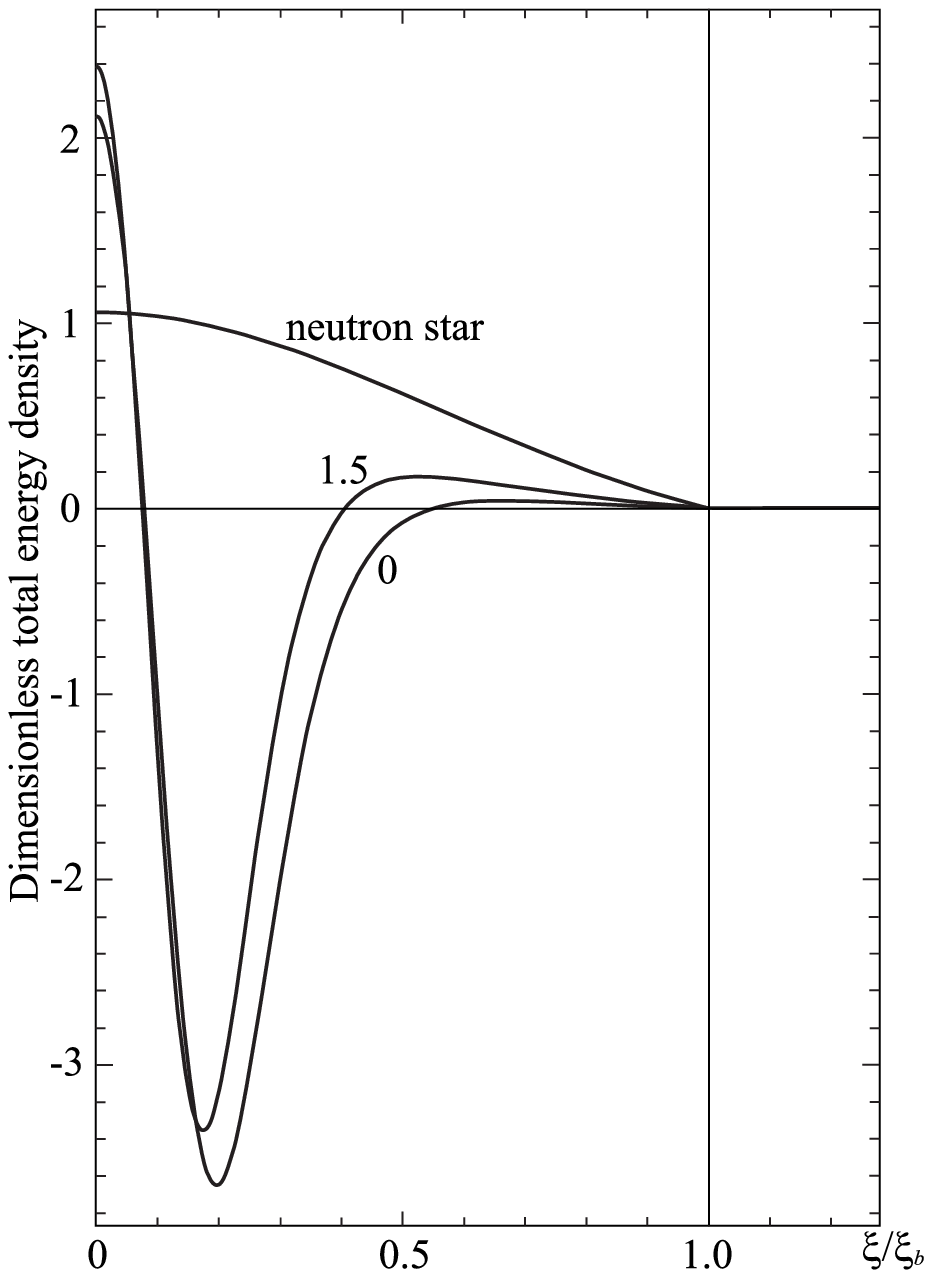}
  \end{center}
\end{minipage}\hfill
\caption{
The fluid energy density $B  (1+\sigma n \theta) \theta^n$ (left panel) and
the total energy density $T_0^0$ (right panel)
from the right-hand side of Eq.~\eqref{Einstein-00_dmls}
(in units of $|V(\varphi(0),\chi(0))|$)  are shown
 as functions of the relative radius $\xi/\xi_b$.
Since the solutions with the wormhole are symmetric with respect to $\xi=0$, the graphs are shown only
for $\xi>0$.
In the left panel, the parameter $\theta_c$ runs the values $0, 0.1, 0.4, 0.7, 1.0, 1.5$, from bottom to top.
In the right panel, all the remaining curves with  $0<\theta_c<1.5$ lie between the presented curves.
The thin vertical line corresponds to the boundary of the fluid.
}
\label{fig_energ_dens}
\end{figure}

\begin{figure}[t]
\centering
  \includegraphics[height=10cm]{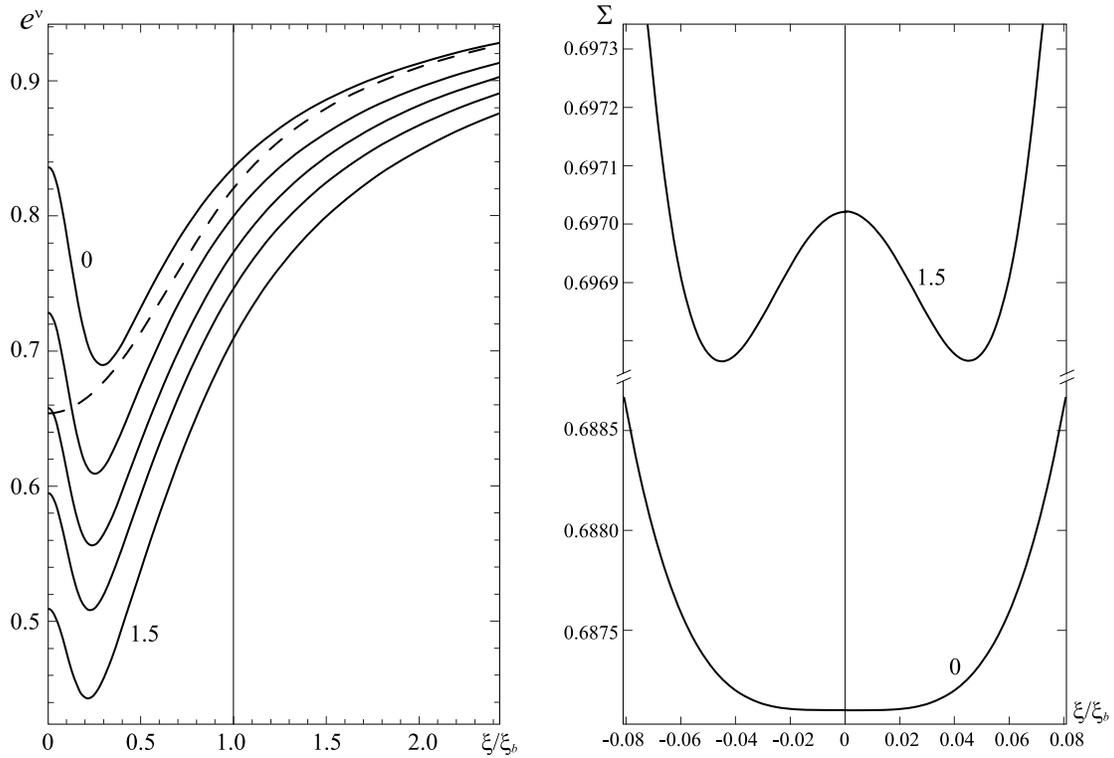}
\caption{The metric functions $g_{tt}=e^{\nu}$ (left panel) and $\Sigma$ (right panel)
 are shown  as functions of the relative radius $\xi/\xi_b$.
In the left panel,
 the thin vertical line corresponds to the boundary of the fluid;
 the dashed line refers to the neutron star;
  for the mixed configurations, the parameter $\theta_c$  runs the values
 $0, 0.4, 0.7, 1.0, 1.5$, from top to bottom.
 Asymptotically, as $\xi\to\pm\infty$, the spacetime is flat
with $\Sigma\to |\xi|$ and $e^{\nu}\to 1$
from  below.
}
\label{fig_metric}
\end{figure}

\begin{figure}[t]
\centering
  \includegraphics[height=6.5cm]{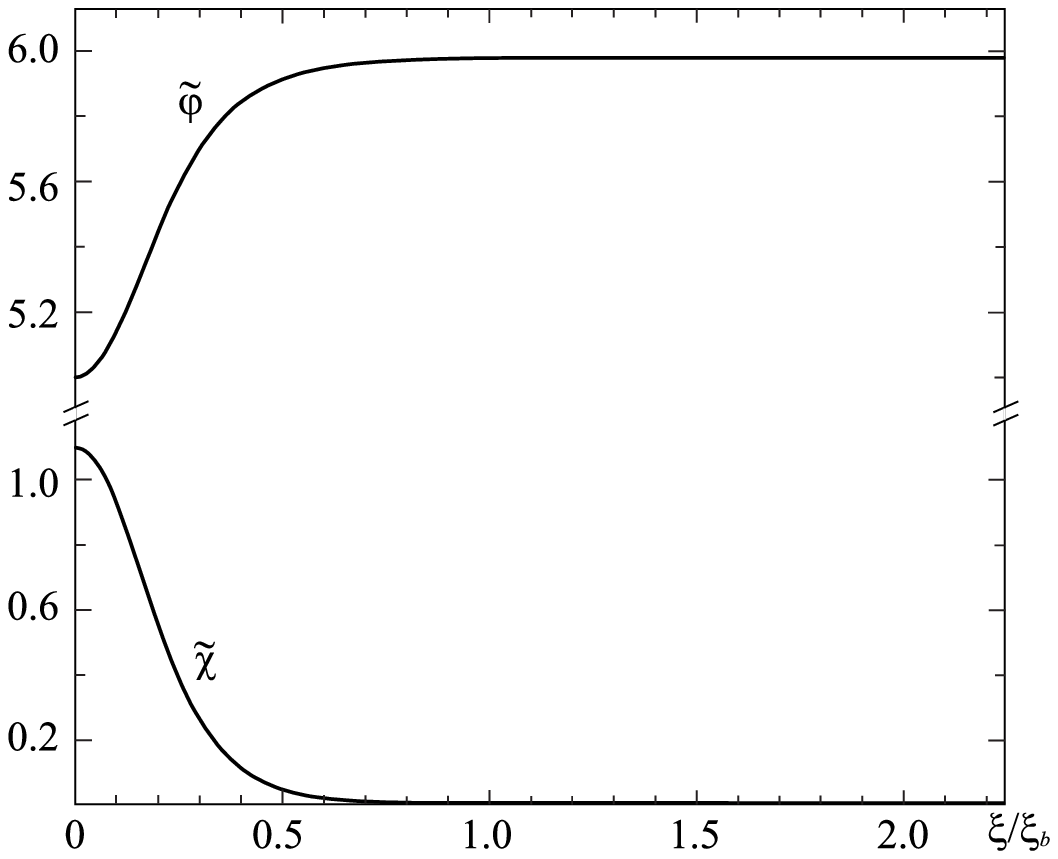}
\caption{The typical behavior of the scalar fields.
 Asymptotically, as $\xi\to\pm\infty$, the field $\tilde \varphi$ tends to $\mu_1$  and
$\tilde\chi$ goes to 0 [see Table~\ref{tab1} and Eqs.~\eqref{asymptotic}-\eqref{asymp_pert}].
}
\label{fig_scal_fields}
\end{figure}

Proceeding in this way, we have obtained the results shown in Figs.~\ref{fig_energ_dens}-\ref{fig_scal_fields}.
The parameters of the system have been chosen so that the masses of the mixed configurations under consideration be of the order of
1-2 solar masses, and their sizes~$\sim 20 \, \text{km}$. These characteristics are typical for neutron stars~\cite{Potekhin:2011xe}.

The results  were obtained by choosing the following fixed values of the parameters:
the characteristic fluid density $\rho_{b c}= 10^{14} \text{g cm}^{-3}$ and its size $L=10\, \text{km}$,
the coupling constants $\Lambda_1=0.1, \Lambda_2=1$, the central values of the scalar fields
$\tilde\varphi_c=5, \tilde\chi_c=1.1$.
This allowed to get configurations with the required characteristics, as shown in Table~\ref{tab1}.

\begin{table}[h]
\caption{Characteristics of the mixed configurations at fixed $\rho_{b c}, L, \Lambda_1, \Lambda_2, \tilde\varphi_c, \tilde\chi_c$ (see in the text).
Here the central  $\rho_{c}=\rho_{b c} \theta_c^n$ and
the maximum  $\rho_{\text{max}}$ densities of the neutron fluid
(both in units of $10^{14} \text{g cm}^{-3}$),
the total mass $M_{\text{tot}}$,
the proper mass $M_{\text{prop}}$ of the neutron matter (both in solar mass units $M_\odot$),
the radius of the neutron fluid $R$ (in kilometers),
and the masses of the scalar fields $\mu_1, \mu_2$ are shown.
For the neutron star with the same values of the polytropic parameters and $\rho_{b c}$, we have
 $M_{\text{tot}}=1.876 M_\odot$ and $R=32.812\, \text{km}$.
 }
\vspace{.3cm}
\begin{tabular}{p{1.5cm}p{1.5cm}p{1.5cm}p{1.5cm}p{1.5cm}p{1.5cm}p{1.5cm}}
\hline \\[-5pt]
$\rho_{c}$ & $\rho_{\text{max}}$  &$ M_{\text{tot}}$ & $ M_{\text{prop}}$& $ R$   & $\mu_1$ & $\mu_2$ \\[2pt]
\hline \\[-7pt]
0.0&0.838&	1.087&	0.526&18.555 & 5.97803 &7.67266\\
0.1 &0.932& 1.173&	0.623&19.201 & 5.97824 &7.67252\\
0.4&1.213&	1.419&	0.902&20.437 & 5.97884 &7.67213\\
0.7&1.489&	1.631&	1.147&20.980 & 5.97940 &7.67175\\
1.0&1.761&	1.805&	1.351&21.132 & 5.97994 &7.67140\\
1.5&2.206&	2.022&	1.613&20.930 & 5.98078 &7.67084\\
\hline
\end{tabular}
\label{tab1}
\end{table}

As seen in Fig.~\ref{fig_scal_fields},
$\tilde\varphi \rightarrow \mu_1$ and $\tilde\chi \rightarrow 0$ at large distances,
i.e., the solutions approach asymptotically the local minimum of the potential \eqref{potph_dmls}.
The arbitrary  constant $\tilde{V}_0$ is chosen so that the value of the potential \eqref{potph_dmls}
be equal to zero at the local minimum, i.e., we take
 $\tilde{V}_0=\Lambda_2 \mu_2^4/4$. Such a choice
 ensures a zero value for the
scalar field energy density as $\xi \rightarrow \pm\infty$, as one can see from Fig.~\ref{fig_energ_dens}.

The asymptotic behavior of the scalar fields can be found in the following form:
\begin{equation}
\label{asymptotic}
    \tilde\varphi=\mu_1-\delta \tilde\varphi, \quad \tilde\chi=\delta \tilde\chi,
\end{equation}
where
$\delta \tilde\varphi, \delta \tilde\chi \ll 1$ behave as
\begin{equation}
    \delta \tilde\varphi \approx  C_{\varphi} \frac{\exp{\left(- \sqrt{2 \Lambda_1 \mu_1^2} \,\, \xi \right)}}{\xi}, \quad
    \delta \tilde\chi \approx  C_{\chi}\frac{\exp{\left(- \sqrt{ (2 \mu_1^2-\Lambda_2 \mu_2^2)} \,\,\xi\right)}}{\xi},
\label{asymp_pert}
\end{equation}
where $C_{\varphi}, C_{\chi}$ are integration constants.

The total masses of the configurations shown in the Table were calculated as follows.
Consider a volume enclosed by a sphere with
circumferential radius $R_c$, corresponding to the center
of the configuration, and another sphere with
circumferential radius $R>R_c$.
Then, using the spherically symmetric metric \eqref{metric_gauss},
the mass $m(r)$ associated with this volume
can  be defined as follows:
\begin{equation}
\label{mass_dm}
m(r)=\frac{c^2}{2 G}R_c+\frac{4\pi}{c^2}\int_{R_c}^{r} T_0^0 R^2   dR.
\end{equation}
For single-throat configurations, the circumferential radius $R_c$ corresponds to the radius
of the wormhole throat defined by $R_c=\text{min}\{R(r)\}$.
However, for the systems considered here, a single-throat configuration will exist only when
$\theta_c=0$. When $\theta_c\neq 0$, as mentioned above,
$R_c$ corresponds to an equator,
while the two throats are located symmetrically away from the center (see in the right panel of Fig.~\ref{fig_metric}).

In the dimensionless variables of Eqs.~\eqref{dimless_var} and \eqref{theta_def}
the expression \eqref{mass_dm} takes the form
\begin{equation}
\label{mass_dmls}
m(\xi)=M^*
\left\{\Sigma_c+
\int_{0}^{\xi} \Big[
B  (1+\sigma n \theta) \theta^n
-\frac{1}{2}\left(\tilde\varphi^{\prime 2}+\tilde\chi^{\prime 2}\right)-\tilde{V}
\Big]\Sigma^2 \frac{d\Sigma}{d\xi'}d\xi'
\right\},
\end{equation}
where the coefficient $M^*$ in front of the curly brackets has the dimension of mass
$$
M^*=\frac{c^3}{2}\sqrt{\frac{B}{8\pi G^3 \rho_{b c}}}.
$$

Note that formally the total mass $M_{\text{tot}}$ is then obtained by taking the upper limit of the integral to infinity,
since the energy density of the scalar fields becomes equal to zero only asymptotically,
as $\Sigma\to \infty$. However, since the scalar fields decay exponentially,
see Eq.~\eqref{asymp_pert},
the integral converges rapidly to its asymptotic value even at small
 $\xi\sim {\cal O}(1)$.
Note also that in evaluating the above integral it is necessary
to perform the calculations in the internal and external regions separately.

One more interesting characteristic of the configurations under consideration is the proper mass $M_{\text{prop}}$
of the neutron matter:
\begin{equation}
\label{mass_nm_prop}
M_{\text{prop}}\equiv m_b N=4\pi \int_0^{r_b} \rho_b R^2 dr=
4\pi \rho_{bc}L^3\int_0^{\xi_b} \theta^n \Sigma^2 d\xi .
\end{equation}
$M_{\text{prop}}$ is equal to the mass
which the baryons of the system would possess altogether,
if they were dispersed throughout a volume so large
that all types of interactions between them could be neglected.
Evaluating the expression (\ref{mass_nm_prop}) for the proper mass,
we can find the number of neutrons $N$ in the system.

From the results obtained, it is possible to delineate the following features of the systems under consideration.
\begin{enumerate}
\item[(1)]
For the values of the parameters used here,
it is seen from the Table that while the central value of the neutron fluid density increases, the masses of the scalar fields
remain practically unchanged.
This corresponds to the fact that the spatial distribution of the scalar fields changes also very slightly for different
amount  of the neutron matter. In this sense, the scalar fields can be regarded as the background ones.
But the total mass of the configuration varies substantially that is obviously caused by a change in the number of neutrons $N$
in the system (see the Table).

\item[(2)] The presence of the ordinary matter results in the following changes in the geometry of the system:
(a)  as the amount of the neutron matter increases,
the redshift function $g_{tt}=e^{\nu}$ increasingly differs from 1
(see Fig.~\ref{fig_metric}), i.e., the system becomes more and more relativistic; (b) at $\theta_c\neq 0$
two weakly marked throats are necessarily present in the system
(see in the right panel of Fig.~\ref{fig_metric}), in contrast to systems supported by
scalar fields only where just one throat is present.

\item[(3)]
One more interesting feature associated with the presence of
the two scalar fields is
a unique shape of the neutron matter distribution along the radius.
The literature in the field offers different systems with an isotropic fluid having a maximum density at the center,
like that illustrated by the curve labeled by ``neutron star'' in Fig.~\ref{fig_energ_dens}.
This applies both to systems with a trivial topology of spacetime (ordinary stars) and to the mixed systems of the ``wormhole plus ordinary matter'' type
 of Refs.~\cite{mix_syst,Dzhunushaliev:2013lna}.
When a system contains an anisotropic fluid, its maximum density may already be located somewhere  between the center
and the edge of the system
(see Refs.~\cite{anis_ord} concerning ordinary stars and
Ref.~\cite{Aringazin:2014rva} where a mixed system is discussed).
For the mixed configurations considered here such a shift of the maximum fluid density takes place even for isotropic matter,
and the central density of the neutron matter can vary from 0 to the values of the order of nuclear density
(see the Table and Fig.~\ref{fig_energ_dens}).
\end{enumerate}

Of course, the obtained quantitative characteristics (masses, sizes, shapes of the neutron matter distribution)
depend essentially on the specific values
of scalar field parameters  (coupling constants, boundary conditions) and of ordinary matter characteristic densities and sizes.
But one might expect that the qualitative behavior of the solutions will remain the same even with a different, physically reasonable choice of the parameters of the system under consideration.

\section{Conclusion}
\label{conclusion}

We have studied equilibrium
mixed configurations consisting of a wormhole
supported by two interacting ghost scalar fields and
threaded by a relativistic polytropic neutron fluid.
The scalar fields play the role of exotic matter
providing a nontrivial topology of spacetime in the system.

Our goal was to investigate the influence that
such a nontrivial topology has
on the distribution of the neutron matter and on physical characteristics of resulting objects.
In doing so,
we compared the obtained mixed configurations with a neutron star modeled by matter with the same EOS.
Having fixed the polytropic parameters and coupling constants of the scalar fields,
we kept track of changes in the distribution of the neutron matter depending on the central value of the neutron matter rest-mass density.

The main results can be summarized as follows:
\begin{enumerate}
\itemsep=-0.2pt
\item[(i)] 
There exist static regular asymptotically flat solutions
describing mixed neutron-star-plus-wormhole systems
in which the neutron matter is concentrated in a finite-size region.
Such configurations may be regarded as
consisting of a neutron star having either a single throat at the center
(at zero central density of the neutron matter)  or an equator
surrounded by a double throat (see Fig.~\ref{fig_metric}).

\item[(ii)]
Due to the presence
of two scalar fields, the neutron fluid always has a maximum density somewhere
between the center of the configuration and the edge of the fluid.
In this case the central density of the neutron matter can vary from 0 to the values of the order of nuclear density (see Table~\ref{tab1} and Fig.~\ref{fig_energ_dens}).
Such a behavior, which is usually characteristic of anisotropic fluids, is attained here even in the case of isotropic fluid.
This is because of the characteristic behavior of the redshift function $g_{tt}=e^{\nu}$ which, in systems with two scalar fields,
has a minimum  shifted away from the center
(see Fig.~\ref{fig_metric}, and also the pure scalar field systems of Ref.~\cite{Dzhunushaliev:2007cs}).

\end{enumerate}

Altogether, as regards the form of the neutron matter distribution, the mixed systems ``two scalar fields plus isotropic matter''
discussed here are similar to the mixed systems ``one scalar field plus anisotropic matter''
of Ref.~\cite{Aringazin:2014rva}, with the main difference related to the behavior
of the redshift function. On the other hand,
the mixed two-field systems considered here differ in principle from the
mixed one-field systems of Refs.~\cite{mix_syst,Dzhunushaliev:2013lna},
for which the maximum  of the neutron fluid density is {\it always} located at the center of a configuration.

In conclusion, we may say a little regarding the question of stability of the considered mixed systems. The results available in the literature indicate
that for pure field configurations (with no ordinary matter)
 wormhole solutions are linearly \cite{Gonzalez:2008wd,Bronnikov:2011if,Bronnikov:2012ch}
and nonlinearly  \cite{Shinkai:2002gv,Gonzalez:2008xk} unstable (see, however,
the recent work \cite{Bronnikov:2013coa} where a method of obtaining
general-relativistic stable wormhole solutions is presented). As shown in our previous studies of mixed ``wormhole plus star'' systems with one scalar field
\cite{Dzhunushaliev:2013lna,Dzhunushaliev:2014mza,Aringazin:2014rva}, 
the presence of ordinary matter did not result in stabilization of the configuration.
One might expect that the inclusion in the system of one more ghost scalar field will not change the stability situation.
Nevertheless,
this issue requires special study, and we plan to perform such a stability analysis later on.

\section*{Acknowledgements}

We gratefully acknowledge support provided by a grant No.~0263/PCF-14 in fundamental research in natural sciences by the Ministry of
Education and Science of Kazakhstan.

\end{document}